\documentstyle[12pt]{article}
\def\appendix#1{
  \addtocounter{section}{1}
  \setcounter{equation}{0}
  \renewcommand{\thesection}{\Alph{section}}
 \section*{Appendix \thesection\protect\indent \parbox[t]{11.715cm} {#1}}
  \addcontentsline{toc}{section}{Appendix \thesection\ \ \ #1}
  }

\newcommand{\newsection}{
\setcounter{equation}{0}
\section}
\def\be{\begin{equation}}
\def\ee{\end{equation}}
\newcommand{\tr}[1]{\:{\rm tr}\,#1}
\def\const{{\rm const}}
\def\e{{\,\rm e}\,}
\def\Evac{E_{{\rm vac}}}

\def\H{H_{{\rm YM}}}
\def\vac{|0\rangle}
\def\d{\partial}
\newcommand{\mn}[1]{\left\langle #1 \right\rangle}
\newcommand{\rf}[1]{(\ref{#1})}
\newcommand{\non}{\nonumber \\*}
\renewcommand{\l}[1]{\left.\frac{\partial }{\partial
\lambda }#1\right|_{\lambda =0}}
\begin{document}
%\begin{titlepage}
\begin{flushright}
ITEP--TH--43/98\\
hep-th/9808189\\
%July, 1998
\end{flushright}
\vspace{1cm}

\begin{center}
{\LARGE Ground state in gluodynamics}
\\[.5cm]
 
{\LARGE and $q\bar{q}$ potential\footnote{
Based on talks given at 10th International Seminar on High Energy Physics 
"Quarks'98", Suzdal, Russia, May 18--24, 1998 and at Workshop on 
"Instantons and Monopoles in the QCD Vacuum", Copenhagen,
Denmark, June 22--27, 1998.}}
\\[.8cm]
%{\LARGE in QCD}\\
%\vspace{1cm}
{\large K.~Zarembo}\\
\vspace{0.5cm}
{\it Department of Physics and Astronomy,}
\\{\it University of British Columbia,}
\\ {\it 6224 Agricultural Road, Vancouver, B.C. Canada V6T 1Z1}
\\ \vskip .2 cm
and\\ \vskip .2cm
{\it Institute of Theoretical and Experimental Physics,}
\\ {\it B. Cheremushkinskaya 25, 117259 Moscow, Russia} \\ \vskip .5 cm
E-mail: {\tt zarembo@itep.ru/@theory.physics.ubc.ca}
\end{center}
%\vskip 2 cm
\begin{abstract}
Static color charges in Yang-Mills theory 
are considered in the Schr\"odinger picture.
Stationary states containing color sources,
interquark potential and confinement criterion
are discussed within the framework based on
integration over gauge transformations which
projects the vacuum wave functional on the
space of physical, gauge-invariant states.
\end{abstract}

%\end{titlepage}

%\setcounter{page}{2}

\newsection{Introduction}

Usually the
potential of interaction between static quark and antiquark
is extracted from Wilson loop expectation value. A Wilson 
loop is essentially non-local operator, since it 
describes the process developing in time. 
On the other hand, static charges are point-like 
 and it might seem that in the Schr\"odinger picture, when 
  stationary states are considered, they can be described by some
local operators. 
Of course, such operators can not be local in terms of the 
original gluon fields, but it is reasonable to look for auxiliary
variables, in which color charge creation operators can be local. The best
candidates for such variables are pure-gauge components of the gluon fields.

Longitudinal gluons 
are unphysical in Yang-Mills theory,
but, once color charges are considered, longitudinal degrees of freedom do not
decouple completely, since the electric field of a static charge is 
longitudinal and longitudinal  gluons
 are responsible for color electric forces. In this respect, it is
useful to introduce gauge degrees of freedom as 
independent variables from the outset.  Such kind of variables naturally 
appears if the gauge invariance of physical states
is imposed by averaging over gauge transformations. An approach 
to the ground state in Yang-Mills theory based on this procedure
was advocated in Refs.~\cite{kk94,dia98}, where it was proposed to
find an approximate vacuum wave functional from variational principle 
applied to the simplest, Guassian variational ansatz.
Averaging over gauge transformations allows to make trial wave 
functional exactly gauge invariant. 
It also
allows to construct easily the
states describing static color sources and to formulate a
relatively simple criterion of confinement in the Hamiltonian framework
\cite{zar98,zar98'}.

\newsection{Vacuum sector}

The ground state in gluodynamics satisfies Schr\"odinger equation
\begin{equation}\label{schr}
\H\Psi =\Evac\Psi ,
\end{equation}
where $\H$ is Yang-Mills Hamiltonian (in $A_0=0$ gauge):
\begin{equation}\label{ham}
\H=\int d^3x\,\left(\frac{1}{2}\,E_i^AE_i^A
+\frac{1}{4}\,F_{ij}^AF_{ij}^A\right).
\end{equation}
When the wave function is a functional of
the gauge potentials, the electric fields act as variational derivatives:
\begin{equation}\label{elect}
E_i^A=-i\,\frac{\delta }{\delta A_i^A}.
\end{equation}
We consider $SU(N)$ gauge group with generators
$T^A$ obeying conventional
normalization conditions
 $\tr T^AT^B=-\delta ^{AB}/2$ and sometimes use matrix
notations for gauge potentials:  $A_i=A_i^AT^A$.

The Hamiltonian \rf{ham} commutes with operators $D_iE_i^A(x)$, and,
apart from the Schr\"o\-din\-ger equation, the
physical states are subject to the Gauss' law constraint:
\begin{equation}\label{gauss}
D_iE_i^A\, \Psi =0.
\end{equation}
%
%\subsection{Projection on gauge-invariant states}
%
The Gauss' law represents in the infinitesimal form the gauge
invariance of the physical states:
\begin{equation}\label{ginv} \Psi [A^\Omega]=\Psi [A],
~~~~~ A_i^\Omega =
\Omega ^{\dagger}\left(A_i+\frac{1}{g}\,\partial
_i\right)\Omega .
\end{equation}
To be more presize, the infinitesimal form of the invariance condition allows
for an additional factor in eq.~\rf{ginv} related to topologically non-trivial
gauge transformations \cite{jac80}. Absence of this factor corresponds to the
case of zero theta-parameter, which is implied hereby. A non-zero theta 
 is discussed in Ref.~\cite{kk94}.

There are many ways to solve the Gauss' law constraint -- to use gauge-invariant
variables \cite{giv}, to fix the 
residual gauge freedom \cite{gf}, or to enforce
gauge invariance projecting the wave function on the physical subspace 
 \cite{kk94,zar98,hmvi98,dia98,zar98'}. The projection
operator has an elegant path-integral representation as an averaging over all
gauge transformations:
\begin{equation}\label{ansatz}
\Psi [A]=\int [DU]\,\e^{-S[A^U]},
\end{equation}
This representation is well known in Yang-Mills theory at finite temperature
\cite{ft,book} and was used to find an approximate
ground state of the Yang-Mills theory by variational method \cite{kk94,dia98}.
In a sense, eq.~\rf{ansatz} is a general solution of the Gauss' law,
since any gauge-invariant functional can be represented in this form by an
appropriate choice of the "bare" state $\exp(-S[A])$.  

\newsection{Static charges}
\subsection{Coupling to heavy fermions}

To introduce static charges, we couple Yang-Mills fields to infinitely heavy
fermions. By infinitely heavy we infer fermions without kinetic term in the
Hamiltonian:
\begin{equation}\label{hamf}
H=\int d^3x\,\left(\frac{1}{2}\,E_i^AE_i^A
+\frac{1}{4}\,F_{ij}^AF_{ij}^A+M\bar{\psi }\psi \right).
\end{equation}
The fermion operators obey anticommutation relations
\begin{equation}\label{ac}
\left\{\psi^{a}_\alpha(x),
\psi ^{\dagger}_{\beta b}(y)\right\}
=\delta _{\alpha \beta }\delta^{a}_{\phantom{a}b}\delta (x-y),
\end{equation}
where $a,b$ and $\alpha ,\beta $ are color and spinor indices,
respectively. The fermion fields transform in the fundamental
representation of $SU(N)$:
\begin{equation}\label{psiu} \psi
^U=U^{\dagger}\psi,~~~~~ \psi ^{\dagger\,U}=\psi^{\dagger}U.
\end{equation}

The ground state of the Hamiltonian \rf{hamf}
can be easily constructed, because
$H$ obeys  simple commutation relations with fermion operators:
\begin{eqnarray}\label{hpsi}
\left[H,\psi _\alpha (x)\right] &=& \pm M\psi _\alpha (x),
~~~~~\left\{
\begin{array}{lr}
+,&\alpha =3,4\\
-,&\alpha =1,2\\
\end{array}\right.
\\*
\left[H,\psi^{\dagger}_\alpha (x)\right] &=&
\pm M\psi^{\dagger}_\alpha  (x),
~~~~~\left\{
\begin{array}{lr}
+,&\alpha =1,2\\
-,&\alpha =3,4\\
\end{array}\right.\,.
\end{eqnarray}
These relations are written in the basis in which $\gamma^0$ is diagonal.
It is clear that 
$\psi _{1,2}(x), \psi ^{\dagger}_{3,4}(x)$ are annihilation
operators and the fermion vacuum is defined by equations
\begin{eqnarray}\label{vacuum}
\psi _{1,2}(x)\vac&=&0,
\non
\psi^{\dagger} _{3,4}(x)\vac&=&0.
\end{eqnarray}
The fermion vacuum is gauge invariant by itself, and the ground state of
the Hamiltonian \rf{hamf} is described by the wave function
\begin{equation}\label{gst}
\Psi _{{\rm vac}}=\Psi [A]\vac=\int [DU]\,\e^{-S[A^U]}\,\vac,
\end{equation}
where $\Psi [A]$ is the vacuum wave functional in gluodynamics -- the
lowest-energy solution of the Schr\"odinger equation \rf{schr}.

Operators $\psi _{3,4}(x), \psi ^{\dagger}_{1,2}(x)$ create excited fermion
states which contain
 static color charges. These states are not gauge invariant
and, thus, it is necessary to project them on the physical subspace. Again,
projection can be realized as averaging over gauge transformations. Therefore,
physical states are obtained by acting of gauge transformed operators
on the fermion vacuum with subsequent integration over the gauge group. This
procedure can be considered as a kind of dressing of the bare fermions
by their static color-electric fields. The dressed operators are gauge-invariant
-- the gauge indices are replaced by global color ones \cite{zar98'}, 
which we mark by prime:
\begin{eqnarray}\label{dressed}
\psi ^{U\,a'}_\alpha(y)&=&
U^{\dagger\,a'}_{\phantom{\dagger\,a'}a}(y)
\psi^a_{\alpha }(y),
\non
\psi ^{\dagger\,U}_{\beta b'}(x)
&=&\psi ^{\dagger}_{\beta b}(x)
U^b_{\phantom{b}b'}(x),
\end{eqnarray}
where $\alpha =3,4$ and $\beta =1,2$ (below spinor indices are omitted
for brevity). 
The simplest example of a color-singlet state containing static charges is
the one with meson quantum numbers:
\begin{equation}\label{mes}
\Psi _M(x,y)=\int [DU]\,\e^{-S[A^U]}\,
\psi^{\dagger\,U}_{a'}(x)
\psi ^{U\,a'}(y)\vac
=\Psi^a_{\phantom{a}b}[A;x,y]\,
\psi^{\dagger}_{a}(x)\psi ^{b}(y)\vac,
\end{equation}
where the gluonic part of the wave function,
\begin{equation}\label{qq}
\Psi^a_{\phantom{a}b}[A;x,y]
=\int [DU]\,\e^{-S[A^U]}\,
U^a_{\phantom{a}a'}(x)U^{\dagger\,a'}_{\phantom{\dagger\,a'}b}(y)
\end{equation}
transforms in the representation
$\bar{N}\otimes N$ of the gauge group
in order to compensate gauge transformations of the fermions.

\subsection{Abelian theory}

The above construction, in fact, generalizes the dressing of electron operators
proposed by Dirac \cite{dir55}: 
\begin{eqnarray}\label{dressedab}
\psi ^{(*)}_\alpha(y)&=&\e^{-ieV(y)}
\psi_{\alpha }(y),
\non
\psi ^{(*)\,\dagger}_{\beta}(x)
&=&\e^{ieV(x)}\psi ^{\dagger}_{\beta}(x),
\end{eqnarray}
, 
where
\begin{equation}\label{defV}
V(x)=\int d^3y\,\frac{1}{-\partial ^2}(x-y)\partial _iA_i(y).
\end{equation}
Dressed operators \rf{dressedab} possess a number of important
properties. They are gauge-invariant and create eigenstates of the free
electro-magnetic Hamiltonian. They also play an important role in QED
eliminating IR
divergencies related to emission
of soft photons \cite{lm95,ir}. Different non-Abelian generalizations of
the operators \rf{dressedab} were proposed \cite{lm95,dr}. 

It is not difficult to reproduce the dressing \rf{dressedab} from the
representation based on averaging over gauge transformations. In Abelian
theory $U=\e^{ie\varphi }$, $A^U_i=A_i+\partial _i\varphi$ and the action
$S[A]$  is
 quadratic; for simplicity we choose the diagonal quadratic form.
Then the
 integration over gauge transformations in \rf{qq} is Gaussian and, for
example, the state \rf{mes} describing two charges of opposite sign
has the form:
\begin{eqnarray}\label{abel}
\Psi [A;x,y]&=&\int [d\varphi ]\,\exp\left[-\frac{1}{2}
\left(A_i+\partial _i\varphi \right)
K\left(A_i+\partial _i\varphi \right)\right]
\e^{ie\varphi (x)}\e^{-ie\varphi (y)}
\non
&=&C\,\e^{ieV(x)}\e^{-ieV(y)}
\exp\left(-\frac{1}{2}A_i^\bot K A_i^\bot
\right),
\end{eqnarray}
where $C$ is a field-independent constant and $A_i^\bot$ denotes
transversal part of the gauge potentials:
\begin{equation}\label{trans}
A_i^\bot=\left(\delta
_{ij}-\frac{\partial _i\partial _j}{\partial ^2}\right)A_j.
\end{equation}
As follows from conformal symmetry, the coefficient function $K$ is equal
to $|p|$ in the momentum space. Then the last factor in \rf{abel} is
nothing but the
vacuum wave functional for free electro-magnetic field. The first two
factors reproduce Dirac dressing operators.
For a generic charged state  
the operator $\psi ^{U}$ will be replaced by $\psi ^{(*)}$ after averaging over
 gauge transformations 
due to the Gaussian nature of the integration over the Abelian gauge group.

\newsection{Interquark potential}

The energy of the state \rf{mes} after obvious subtractions determines
$q\bar{q}$ interaction potential:
\begin{equation}\label{e12}
\frac{\left\langle\Psi_{M}|H|\Psi_{M}\right\rangle}
{\left\langle\Psi_{M} |\Psi_{M} \right\rangle}
=2M+\Evac+V(x-y).
\end{equation}
Matrix elements of gauge-invariant states, like those entering
eq.~\rf{e12}, are proportional to the volume of the gauge group. The
representation \rf{ansatz} allows to get rid of this infinite factor
easily \cite{kk94}.  For example, the norm of the vacuum state is
\begin{equation}\label{norm}
\left\langle\Psi |\Psi \right\rangle=\int
[DU][DU'][dA]\, \e^{-S[A^U]-S[A^{U'}]}=
\const\,\int [DU][dA]\,\e^{-S[A^U]-S[A]}.
\end{equation}
This expression can be viewed as a partition function of a three-dimensional
statistical system. Matrix elements of the Hamiltonian are determined by a
linear response of this system on perturbation by the operator
\begin{equation}\label{defr}
R[A]=\int d^3x\,\left(\frac{1}{2}\,\frac{\delta
^2S[A]}{\delta A_i^A\delta
A_i^A}-\frac{1}{2}\,\frac{\delta S[A]}{\delta A_i^A}
\,\frac{\delta S[A]}{\delta A_i^A}
+\frac{1}{4}\,F_{ij}^AF_{ij}^A\right),
\end{equation}
which, essentially, is the Hamiltonian. So, defining the partition function
\begin{equation}\label{defz}
Z=\int
[DU][dA]\,\e^{-S[A^U]-S[A]+\frac{\lambda }{2}\,R[A^U]+\frac{\lambda
}{2}\,R[A]}
\end{equation}
we express, for example,  the vacuum energy in gluodynamics 
as a derivative of the
free energy of the statistical system \rf{defz} with respect 
to $\lambda$:
\begin{equation}\label{evac}
\Evac=\frac{\left\langle\Psi |H_{\rm YM}|\Psi \right\rangle}
{\left\langle\Psi |\Psi \right\rangle}=\l{\ln Z}.
\end{equation}
This quantity is proportional to the volume and contains UV divergent
contribution from zero-point oscillations. Regularized energy density can
be related to gluon condensate \cite{SVZ79}:
\begin{equation}\label{edens}
\Evac=\frac{\beta (\alpha _s)}{16\alpha _s}\left\langle
0\left|F_{\mu \nu }^AF^{A\mu \nu}\right|0\right\rangle
\,{\rm Vol}+\mbox{UV divergent terms},
\end{equation}
where $\beta (\alpha _s)$ is the $\beta $-function.

The quark-antiquark potential can be represented 
in the form \cite{zar98,zar98'}:
\begin{equation}\label{v12}
V(x-y)=\l{\ln\left\langle\tr U(x)U^{\dagger}(y)\right\rangle},
\end{equation}
where averaging is defined by eq.~\rf{defz}. 
The confining behavior of the 
$q\bar{q}$ potential is rather natural from the point of view
of this representation. Really, 
the correlation function in eq.~\rf{v12} 
must decrease at infinity. 
So, its logarithm, the derivative of which determines the
 potential, always increases. This does not mean that the
potential is always rising at infinity. If
the correlator of gauge rotations increases as a power of distance with
fixed exponent, the potential decreases. 
However, if a mass gap is generated and the correlator falls exponentially,
the potential grows linearly:
 \begin{equation}\label{expfall}
\left\langle\tr U(x)U^{\dagger}(y)\right\rangle
\propto\e^{-mr},~~~r=|x-y|\rightarrow\infty,
\end{equation}
\begin{equation}\label{Vqq}
V(r)=\sigma r+\ldots,
\end{equation}
with the string tension determined by the derivative of the mass
with respect to $\lambda$:
\begin{equation}\label{strten}
\sigma =-\left.\frac{\partial m}{\partial \lambda}\right|_{\lambda =0}.
\end{equation}
Hence, the confinement arises due to generation of a mass gap in the
averaging over gauge transformations.
More presizely, the confining
potential is generated if there is a nonzero linear response of the mass
gap on the perturbation by the Hamiltonian in eq.~\rf{defz}.

\subsection{Short-distance behavior}

At short distances the potential can be calculated perturbatively.
The free-gluon (zeroth order perturbative) wave functional is Gaussian
 -- the action $S$ is quadratic:
\be\label{quad}
S[A]=\frac12\,A_i^AKA_i^A+\ldots.
\ee
We do not specify the form of the kernel $K$, since it drops from the 
final answer. The matrices of gauge transformations are expanded as
$U=\e^{g\Phi}=1+g\Phi^AT^A+\ldots$, and $A_i^U=A_i+\partial_i\Phi+\ldots$. 
The gauge potentials can be integrated out in 
this approximation by the change of
variables: $A_i=\bar{A}_i-\d_i \Phi/2$, and the partition function 
\rf{defz} takes the form
\be
Z=\const\,\int[d\Phi]\,\exp\left[-\frac14\,\d_i\Phi^A
\left(K+\frac{\lambda}{2}\,K^2\right)\d_i\Phi^A\right].
\ee
For the correlator which defines the interquark potential we get:
\begin{eqnarray}
\left\langle\tr U(x)U^{\dagger}(y)\right\rangle
&=&N+\frac{g^2}{2}\mn{\Phi^A(x)\Phi^A(y)
-\frac12\,\Phi^A(x)\Phi^A(x)
\right.\non && \left. -\frac12\,\Phi^A(y)\Phi^A(y)}+O(g^4)
\non \nonumber &=&
N+\frac{g^2(N^2-1)}{2}\Bigl(D(x-y)-D(0)\Bigr)
+O(g^4),
\end{eqnarray}
where $D$
is the propagator of the field $\Phi$: $D^{-1}=-\d^2(K+
\lambda K^2/2)/2$. The equation \rf{v12} now yields the Coulomb
potential with correct coefficient:
\begin{eqnarray}\label{1/r}
V(x-y)&=&-\frac{g^2(N^2-1)}{2N}
\left(\frac{1}{-\partial ^2}(x-y)
-\frac{1}{-\partial ^2}(0)\right)
\non &=&-\frac{g^2(N^2-1)}{8N\pi r}
+\mbox {self-energy}.
\end{eqnarray}

\subsection{Large-distance behavior}

The large-distance asymptotics of the potential is determined by
infrared structure of the vacuum wave functional, which, generally
speaking, is unknown. For illustrative  purposes we present here a
calculation of the string tension under particular assumptions
about the dependence of the wave functional on long-wavelength fields
\cite{zar98}. 

The starting point is derivative expansion, 
which probably is justified at large distances.
The main assumption is that the derivative expansion of the kernel $K$
in eq.~\rf{quad} starts from the constant term (in momentum 
representation): $K(p)=M+O(p^2)$. This assumption is compatible with
variational estimates of Ref.~\cite{kk94}; however, arguments in favor of
$M=0$ also exist \cite{Gre79}. If we accept that $M\neq 0$, cubic and
higher terms in $S[A]$, as well as the magnetic part of the
Hamiltonian, are
irrelevant being of higher order in derivatives.
Thus, the action in \rf{defz} takes the form of a local expression
quadratic in gauge fields:
\be
Z=\int [dA]\,[DU]\,\exp\left[-\frac{1}{2}
\left(M+\frac{\lambda}{2}\,M^2\right) \int d^3x\,\left(A_i^AA_i^A
+A_i^{U\,A}A_i^{U\,A}\right)
\right].
\ee
The integration over gauge potentials can be eliminated by the change 
of variables $A_i=\bar{A}_i-\partial _iUU^{\dagger}/2g$. The remaining 
path integral over gauge
transformations is the partition function of three-dimensional 
principal chiral field with local action:
\be
Z=\int [dA]\,[DU]\,\exp\left[-\frac{1}{2g^2}
\left(M+\frac{\lambda}{2}\,M^2\right) \int d^3x\,
\tr\partial _iU^{\dagger}\partial _iU
\right].
\ee 

Principal chiral field is expected to produce a mass gap \cite{book}. 
The simplest way 
to see it is to consider $U$ and $U^{\dagger}$ as independent
complex matrices and to impose the unitarity condition introducing a 
Lagrange multiplyer. If the Lagrange multiplyer acquires non-zero
vacuum expectation value, the field $U$ has massive propagator.
Following Ref.~\cite{kk94}, we use mean field approximation to
determine the mass gap. The gap equation follows from unitarity condition
$\mn{UU^{\dagger}}={\bf 1}$:
\begin{equation}\label{gap}
\int\frac{d^3p}{(2\pi)^3}\,\frac{1}{p^2+m^2}=
\frac{1}{2g^2N}\left(M+\frac{\lambda}{2}\,M^2\right).
\end{equation}
Differentiating this equation in $\lambda$ we get, according to
\rf{strten},
\be
\sigma=-\frac{\partial m}{\partial \lambda}=\frac{\pi M^2}{g^2N}
=\frac{M^2}{4\alpha_sN}.
\end{equation}

This relation is obtained within the effective low-energy theory. Hence, the
coupling here is not the running one; it is fixed on the typical
energy scale:
$\alpha_s\equiv\alpha_s(M)$. It is worth mentioning that the corrections to
the zeroth order of derivative expansion and to the mean field approximation
are parametrically not suppressed and can only be numerically small 
\cite{kk94}. 

\newsection{Discussion}

The averaging over gauge transformations appears to be very convenient
for consideration of static color charges in the Schr\"odinger 
picture. It allows to formulate rather simple criterion of confinement.
But originally integration over the gauge group was proposed in the context
of the variational approach \cite{kk94}.
In principle, variational calculations can be done analytically for
the simplest Gaussian ansatz projected on the space of gauge-invariant
states. The main problem in this approach is correct UV renormalization.
If UV divergencies are not removed by standard counterterms, a variational
approximation looses sense. The UV structure of the vacuum wave functional
was discussed from different point of views in Refs.~\cite{ren}.

Another interesting question concerns topologically 
non-trivial gauge transformations. Kinematical consequences of
large gauge transformations are well known \cite{jac80}. The
dynamics of topologically non-trivial configurations of the field $U$ was
considered recently \cite{bgkk98} for the Gaussian variational
ansatz averaged over the gauge group \cite{bgkk98}. The skyrmion in
this model was interpreted by the authors of Ref.~\cite{bgkk98}
as a Hamiltonian counterpart of the instanton.

\subsection*{Acknowledgments}

The author is grateful to D.~Diakonov,
I.~Kogan, Y.~Makeenko,
M.~Polikarpov, G.~Semenoff and A. Zhitnitsky for
discussions. This work was supported by NATO Science Fellowship and, in
 part, by CRDF grant 96-RP1-253, INTAS grant 96-0524, RFFI grant
 97-02-17927 and grant 96-15-96455 of the support of scientific schools.


\begin{thebibliography}{99}
\addtolength{\itemsep}{-6pt}

\bibitem{kk94}
I.I.~Kogan and A.~Kovner, Phys. Rev. D52 (1995) 3719, hep-th/9408081.

\bibitem{dia98}
D. Diakonov, hep-th/9805137.

\bibitem{zar98}
K. Zarembo, Phys. Lett. B421 (1998) 325, hep-th/9710235.

\bibitem{zar98'}
K. Zarembo, hep-th/9806150.

\bibitem{jac80}
R. Jackiw, Rev. Mod. Phys. 52 (1980) 661.

\bibitem{giv}
J. Goldstone and R. Jackiw, Phys. Lett. B74 (1978) 81;\\
Yu.A. Simonov, Sov. J. Nucl. Phys. 41 (1985) 835 
[Yad. Fiz. 41 (1985) 1311];\\ 
M. Bauer, D.Z. Freedman and P.E. Haagensen, Nucl. Phys. B428 (1994) 147,
hep-th/9405028;\\
P. E. Haagensen and K. Johnson, Nucl. Phys. B439 (1995) 597, 
hep-th/9408164;\\
D. Karabali and V.P. Nair, Nucl. Phys. B464 (1996) 135, 
hep-th/9510157;\\ 
P. E. Haagensen, K. Johnson and C. S. Lam, Nucl. Phys. B477 (1996) 273,
 hep-th/9511226;\\
 D. Karabali, C. Kim and V. P. Nair, hep-th/9804132.

\bibitem{gf}
N.H. Christ and T.D. Lee, Phys. Rev. D22 (1980) 939;\\
P. van Baal, hep-th/9511119; hep-th/9711070.

\bibitem{hmvi98}
C. Heinemann, C. Martin, D. Vautherin and E. Iancu, hep-th/9802036.

\bibitem{ft}
A.M.~Polyakov, Phys. Lett. B72 (1978) 477;\\
L.~Susskind, Phys. Rev. D20 (1979) 2610;\\
B.~Svetitsky, Phys. Rep. 132 (1986) 1.

\bibitem{book}
A.M.~Polyakov, 
{\it Gauge Fields and Strings} (Harwood Academic
Publishers, 1987).

\bibitem{dir55}
P.A.M. Dirac, Can. J. Phys. 33 (1955) 650; {\it The Principles of Quantum
Mechanics} (Oxford, Clarendon Press, 1958).

\bibitem{lm95}
M. Lavelle and D. McMullan, Phys. Rep. 279 (1997) 1, hep-ph/9509344.

\bibitem{ir}
E. Bagan, M. Lavelle and D. McMullan, Phys. Rev. D56 (1997) 3732, 
hep-th/9602083; D57 (1998) 4521, hep-th/9712080. 

\bibitem{dr}
P.E. Haagensen and K. Johnson, hep-th/9702204;\\
M. Lavelle and D. McMullan, hep-th/9805013.
\bibitem{SVZ79}
M.~Shifman, A.~Vainshtein and V.~Zakharov, Nucl. Phys. B147 (1979)
385.
\bibitem{Gre79}
J.~Greensite, Nucl. Phys. B158 (1979) 469; Phys.~Lett. B191 (1987)
431;\\
J.~Greensite and J.~Iwasaki, Phys. Lett. B223 (1989) 207.

\bibitem{ren}
W.E.~Brown and I.I.~Kogan, hep-th/9705136;\\
K. Zarembo, Mod. Phys. Lett. A13 (1998) 1709, hep-th/9803237; 
1795, hep-ph/9804276.

\bibitem{bgkk98}
W. Brown, J.P. Garrahan, I.I. Kogan and A. Kovner, 
hep-ph/9808216. 

\end{thebibliography}
\end{document}